\documentclass[12pt]{article}
\usepackage{amsmath}
\usepackage{amssymb}
\usepackage{slashed}
\usepackage[square,numbers,sectionbib,compress]{natbib}
\usepackage{graphicx}
\usepackage{breqn}
\usepackage{bbold}
\usepackage{xcolor}
\usepackage{float}
\usepackage{graphics, setspace}
\usepackage{framed}
\usepackage[export]{adjustbox}
\usepackage[colorlinks=true,urlcolor=black,anchorcolor=black,citecolor=black,filecolor=black,linkcolor=black,menucolor=black,linktocpage=true,pdfproducer=medialab,pdfa=true]{hyperref}
\allowdisplaybreaks

\newcommand{\eps}{\varepsilon}

\newcommand{\Op}{\mathcal{O}}

\newcommand\deldotPPOW[2]       {(\Delta\cdot P_{#1})^{#2}}

\newcommand\parder[1]       {\partial^{#1}}
\newcommand\SUM[2]          {\sum_{#1}^{#2}}

\usepackage{pslatex}
\usepackage[latin1]{inputenc}
\usepackage[T1]{fontenc}

\makeatletter
\DeclareRobustCommand{\cev}[1]{%
  {\mathpalette\do@cev{#1}}%
}
\newcommand{\do@cev}[2]{%
  \vbox{\offinterlineskip
    \sbox\z@{$\m@th#1 x$}%
    \ialign{##\cr
      \hidewidth\reflectbox{$\m@th#1\vec{}\mkern4mu$}\hidewidth\cr
      \noalign{\kern-\ht\z@}
      $\m@th#1#2$\cr
    }%
  }%
}
\makeatother

\begin{document}

\numberwithin{equation}{section}

\begin{titlepage}
\noindent
\hfill June 2024\\
\vspace{0.6cm}
\begin{center}
{\LARGE \bf 
    All-order Feynman rules for leading-twist gauge-invariant operators in QCD}\\ 
\vspace{1.4cm}

\large
G.~Somogyi$^{\, a}$ and S.~Van Thurenhout$^{\, a}$\\
\vspace{1.4cm}
\normalsize
{\it $^{\, a}$HUN-REN Wigner Research Centre for Physics, Konkoly-Thege Mikl\'os u. 29-33, 1121 Budapest, Hungary}\\
\vspace{1.4cm}

{\large \bf Abstract}
\vspace{-0.2cm}
\end{center}
We present the Feynman rules for leading-twist gauge-invariant quark and gluon operators with an arbitrary number of total derivatives and applicable to any order in perturbation theory. This generalizes previous results and constitutes a necessary ingredient in the computation of the matrix elements of the corresponding operators. The results are written in a form appropriate for implementation in a computer algebra system. To illustrate the latter we provide implementations in {\tt Mathematica} and \textit{{\tt FORM}}, which are made available at \url{https://github.com/vtsam/NKLO}.
\vspace*{0.3cm}
\end{titlepage}

\section{Introduction}
\label{sec:theory}
Composite operators play an important role in the study of the strong interaction and hadronic structure. For example, the hadronic matrix elements of such operators give rise to various types of parton distribution functions. In inclusive processes such as deep-inelastic scattering, one can probe the forward operator matrix elements (OMEs), which define the forward parton distributions like the standard PDFs. Exclusive processes on the other hand are sensitive to the non-forward OMEs, meaning there is a non-zero momentum flow through the operator vertex. Such matrix elements then define the non-forward parton distributions. Standard examples of the latter are the generalized parton distribution functions~\cite{Muller:1994ses,Ji:1996ek,Ji:1996nm,Radyushkin:1996nd,Radyushkin:1996ru,Diehl:2003ny}, which can be accessed experimentally from e.g.~deeply-virtual Compton scattering. Their study will be a major goal of a future electron-ion collider~\cite{Boer:2011fh,AbdulKhalek:2021gbh}.\newline

Besides knowledge about the distributions themselves, phenomenological studies also require information about their scale dependence. The latter is set by the anomalous dimension of the operator defining the distribution and can be computed perturbatively in the strong coupling $\alpha_s$ by renormalizing the partonic matrix elements of the operator. Of course, in order to perform such computations the Feynman rules of the appropriate operator vertices are necessary. In forward kinematics, i.e. when there is no momentum flowing through the operator vertex, the computation and renormalization of the necessary matrix elements is, at least in principle, straightforward for the flavor-non-singlet sector. The analysis of the flavor-singlet one is more complicated however due to the necessity of taking into account non-gauge-invariant (alien) operators, see e.g.~\cite{Hamberg:1991qt,Matiounine:1998ky,Blumlein:2022ndg,Falcioni:2022fdm,Gehrmann:2023ksf,Falcioni:2024xyt}. The Feynman rules for the necessary operator vertices are known to order $g_s^4$ with $\alpha_s=g_s^2/(4\pi)$, allowing for the computation of the matrix elements up to the four-loop level, see e.g.~\cite{Falcioni:2022fdm,Gehrmann:2023ksf,Floratos:1977au,Floratos:1978ny,Mertig:1995ny,Kumano:1997qp,Hayashigaki:1997dn,Bierenbaum:2009mv,Klein:2009ig,Blumlein:2001ca,Velizhanin:2011es,Velizhanin:2014fua,Moch:2017uml,Moch:2021qrk,Falcioni:2023luc,Falcioni:2023vqq,Falcioni:2023tzp,Moch:2023tdj,Gehrmann:2023iah,Kniehl:2023bbk} (and references therein). However in the case of non-forward kinematics, which is relevant in the study of exclusive processes, the momentum flowing through the operator vertex is non-zero. This implies that, during the renormalization procedure, one has to take into account mixing with total-derivative operators. Some direct computations of the non-forward matrix elements of the local operators, and their corresponding renormalization, were performed e.g.~in \cite{Gracey:2009da,Kniehl:2020nhw,Moch:2021cdq,VanThurenhout:2022nmx}. However, the Feynman rules associated to total-derivative operators were not explicitly given. This will be the subject of the current article. In particular, we will determine expressions for the Feynman rules for the operator vertices with an arbitrary number of total derivatives and an arbitrary number of gluons. The latter implies that our expressions are applicable to any order in perturbation theory. We focus our attention on the determination of the Feynman rules for quark and gluon operators in the leading-twist approximation, which corresponds to keeping only the leading term in an expansion in inverse powers of some hard scale of the scattering process. While subleading-twist corrections are becoming increasingly important, see e.g.~\cite{Anselmino:1994gn,Liang:2000gz,Barone:2001sp,Braun:2009mi,Braun:2022gzl}, we do not consider them in this work. \newline

The paper is organized as follows. In the Sec.~\ref{sec:setup} we set up the notation and summarize our conventions. The next section then steps through the derivation of the Feynman rules. Instead of immediately jumping to the result with an arbitrary number of total derivatives and gluons, we first explain the derivation of some simpler operator vertices. The final result is relatively straightforward to implement in a computer algebra language. As illustration, we provide both a {\tt Mathematica} and a \textit{{\tt FORM}}~\cite{Vermaseren:2000nd,Kuipers:2012rf} implementation, which are made available at \url{https://github.com/vtsam/NKLO}. The use of these codes is briefly explained in Sec.~\ref{sec:implementation}. Conclusions and an outlook are provided in Sec.~\ref{sec:conclusion}.

\section{Setup and conventions}
\label{sec:setup}
We consider the leading-twist spin-$N$ quark operators in QCD which are generically of the following form~\footnote{This definition is appropriate for flavor-singlet operators. The flavor-non-singlet ones have an extra factor of $\lambda^{\alpha}$, representing the generators of the flavor group. In the following we omit this factor. Furthermore, we will also not explicitly write the fundamental indices of the quark fields.}
\begin{equation}
\label{eq:operators}
    \Op_{k,0,N-k-1} = \mathcal{S}\partial_{\mu_1}\dots\partial_{\mu_k}[\overline{\psi}\Gamma D_{\rho_{1}}\dots D_{\rho_{N-k-1}} \psi]
\end{equation}
which is a special case of the more general
\begin{equation}
\label{eq:generalOP}
    \Op_{p,q,r} = \mathcal{S}\partial_{\mu_1}\dots\partial_{\mu_p}[(D_{\nu_{1}}\dots D_{\nu_{q}}\overline{\psi})\Gamma (D_{\rho_{1}}\dots D_{\rho_{r}} \psi)].
\end{equation}
Here $\psi$ represents the quark field, $D_{\mu}$ is the QCD covariant derivative and $\Gamma$ denotes a generic Dirac gamma matrix. For phenomenological studies, the following choices for the latter are relevant:
\begin{itemize}
    \item $\Gamma=\gamma_{\rho}$ for the description of unpolarized scattering,
    \item $\Gamma=\gamma_{5}\gamma_{\rho}$ for the description of polarized scattering and
    \item $\Gamma=\sigma_{\mu\rho}\equiv \frac{1}{2}[\gamma_{\mu},\gamma_{\rho}]$ for the study of transverse hadronic structure.
\end{itemize} 
In the leading-twist approximation, the Lorentz indices are symmetrized and traces are subtracted, which is denoted by $\mathcal{S}$. In practice, this operation is implemented by contracting the operator and its matrix elements with $N$ copies of an arbitrary lightlike vector, $\Delta^2=0$. This implies that the unpolarized and polarized matrix elements are now Lorentz scalars. However, the transverse matrix elements have one free Lorentz index left as only the $\rho$ index of $\sigma_{\mu\rho}$ is contracted with a $\Delta$. The remaining Lorentz index $\mu$ can be removed, if desired, by the introduction of another auxiliary vector.\newline

In principle the Feynman rules for the operators in Eq.~(\ref{eq:operators}) are derived from the path integral formulation of QCD. However, in practice it turns out that this is not necessary. Instead, one can simply expand the covariant derivatives and replace any field hit by an ordinary derivative by the momentum flowing through this field. Formally this of course corresponds to performing the Fourier transform to momentum space. In this procedure one has to take care of possible sign conventions. Generically, there are two possible origins for extra signs:
\begin{enumerate}
    \item \textbf{Conventions for the sign of the strong coupling in the covariant derivative.} One either has
    \begin{align}
        &D_{\mu}\psi = \partial_{\mu}\psi-i g_s A_{\mu}\psi\\
        &D_{\mu}\overline{\psi} = \partial_{\mu}\overline{\psi}+i g_s A_{\mu}\overline{\psi}
    \end{align}
    or
    \begin{align}
        &D_{\mu}\psi = \partial_{\mu}\psi+i g_s A_{\mu}\psi\\
        &D_{\mu}\overline{\psi} = \partial_{\mu}\overline{\psi}-i g_s A_{\mu}\overline{\psi}.
    \end{align}
    Here $A_{\mu} = t^{c}A_{\mu,c}$ with $t^{c}$ the generators of the color group.
    \item \textbf{Conventions for the momentum routing.} When the quark field is hit by a derivative, its momentum assignment is
    \begin{align}
        &\partial_{\mu}\psi(p) \rightarrow - i p_{\mu}\psi\\
        &\partial_{\mu}\overline{\psi}(p) \rightarrow + i p_{\mu}\overline{\psi}.
    \end{align}
    This, however, assumes the momentum and particle flow to point in the \underline{same direction}, i.e.~\footnote{All Feynman diagrams in this work are drawn using \textit{FeynGame}~\cite{Harlander:2020cyh,Harlander:2024qbn}.}
\begin{figure}[H]
\centering
\includegraphics[width=0.4\textwidth, trim = 72 536 72 72]{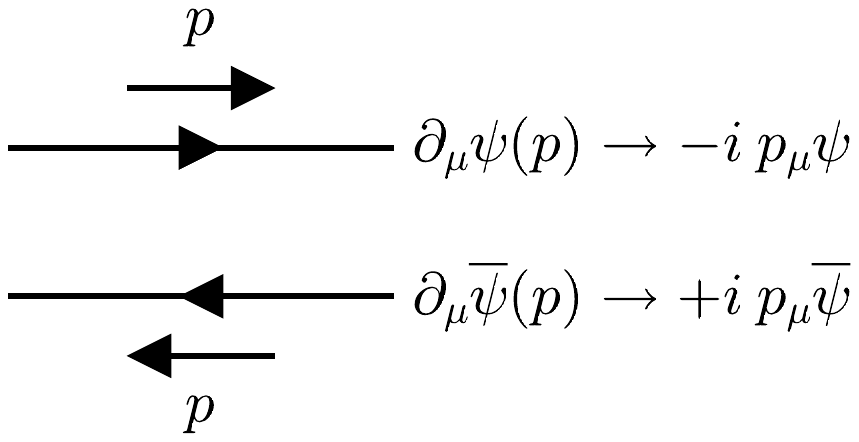}
\end{figure}
If instead the momentum flows in the \underline{opposite direction} to the particle flow, there is a sign flip, i.e.
\begin{figure}[H]
\centering
\includegraphics[width=0.4\textwidth, trim = 72 536 72 72]{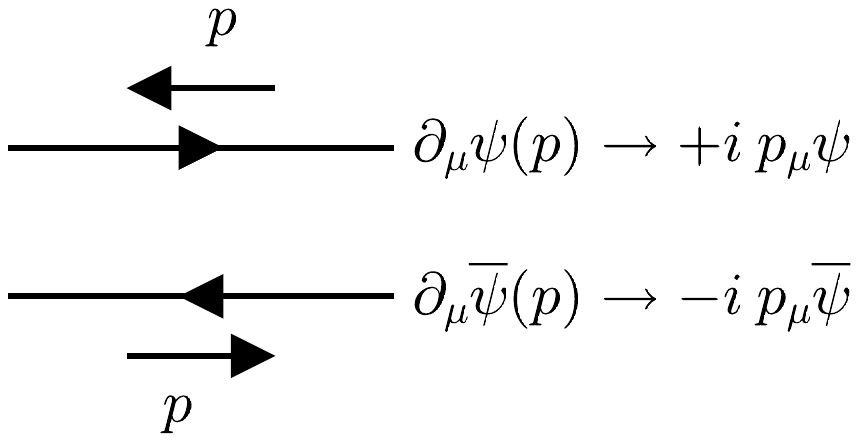}
\end{figure}

\end{enumerate}
Generically now, the operator vertices to be considered have the following form~\footnote{Here and in the following, we do not draw the momentum arrows explicitly as we determine the Feynman rules for arbitrary momentum flows.}
\begin{figure}[H]
\centering
\includegraphics[width=0.5\textwidth,trim = 72 515 72 72]{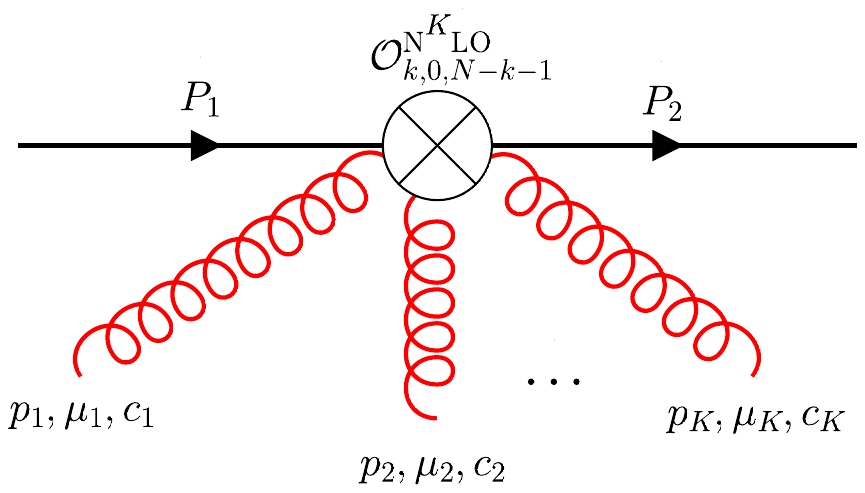}
\end{figure}
meaning our notation is as follows
\begin{itemize}
    \item The momenta of the fermionic fields in the operator are assigned as $\overline{\psi}(P_2)$ and $\psi(P_1)$ respectively.
    \item When computing the matrix elements of the operators at the $L$-loop level, one needs to consider operator vertices with up to $L$ gluons attached, which follows from the expansion of the covariant derivatives in Eq.~(\ref{eq:operators}). Generically we will denote the number of gluons of the operator vertex by $K$. The gluon field with index $g\in [1,K]$ then has momentum $p_g$, Lorentz index $\mu_g$ and a color index $c_g$. All gluon momenta will always be taken to be incoming.
    \item The number of total derivatives in the operator is denoted by $k$ which, for some spin-$N$ operator, lives in the range $k\in[0,N-K-1]$.
\end{itemize}
For the purpose of this work we will use dynamical conventions for easy adaptability to one's own preferences. In particular we set
\begin{align}
\label{eq:conventionsD1}
    &D_{\mu}\psi = \partial_{\mu}\psi- \delta_D(i g_s  A_{\mu}\psi), \\ \label{eq:conventionsD2}
    &D_{\mu}\overline{\psi} = \partial_{\mu}\psi+ \delta_D(i g_s A_{\mu}\overline{\psi}),\\ \label{eq:conventionsp1}
    &\partial_{\mu}\psi(P_1)\rightarrow \delta_2 (i\: P_{1\mu}),\\ \label{eq:conventionsp2}
    &\partial_{\mu}\overline{\psi}(P_2)\rightarrow \delta_1 (i\: P_{2\mu})
\end{align}
with $\delta_D,\delta_1,\delta_2=\pm 1$. The values of $\delta_{1,2}$ are related to the choice of momentum routing. One popular choice is to take all momenta to be flowing into the operator vertex, in which case $\delta_1=\delta_2=-1$. Another often-used option is the so-called physical momentum routing, in which case the momenta and particle flows are chosen to be parallel, setting $\delta_1=-1, \delta_2=+1$.\newline

An important cross-check for the Feynman rules of total-derivative operators is that they should vanish in the forward limit, i.e. when the momentum flowing through the operator vertex is zero. For the choices of momentum routing described above, this limit corresponds to setting
\begin{equation}
    \begin{cases}
        P_1+P_2+\sum\limits_{i=1}^{K}p_i = 0 &\text{[all momenta incoming]} \text{\: or}\\
        P_2=P_1+\sum\limits_{i=1}^{K}p_i  &\text{[physical momentum routing]}.
    \end{cases}
\end{equation}
Finally, when one arrives at an expression for the operator vertex, one can perform several cross-checks to ensure the correctness of the result:
\begin{itemize}
    \item The overall momentum scaling should be $p^{N-K-1}$.
    \item For fixed values of $N$ and $k$, the result should be a polynomial in the momenta.
    \item Total-derivative operators do not contribute in the forward limit. As such, the Feynman rule associated to $\Op_{k,0,N-k-1}$ with $k>0$ is expected to vanish in this limit.
    \item Finally, several expressions for the operator vertices are already known, especially for $k=0$. Hence one should also cross-check against such previous results. The Feynman rules for operators with up to four gluons can be found e.g.~in \cite{Falcioni:2022fdm,Gehrmann:2023ksf,Floratos:1977au,Floratos:1978ny,Mertig:1995ny,Kumano:1997qp,Hayashigaki:1997dn,Bierenbaum:2009mv,Klein:2009ig,Blumlein:2001ca,Velizhanin:2011es,Velizhanin:2014fua,Moch:2017uml,Moch:2021qrk,Falcioni:2023luc,Falcioni:2023vqq,Falcioni:2023tzp,Moch:2023tdj,Gehrmann:2023iah,Kniehl:2023bbk} and references therein. To be specific, we will compare our results against the rules presented in \cite{Gehrmann:2023ksf} and \cite{Klein:2009ig}. Both works present the quark operator rules with up to three additional gluons. However, \cite{Gehrmann:2023ksf} assumes all momenta to be incoming while \cite{Klein:2009ig} uses the physical momentum routing. Hence cross-checking our results against both provides a strong check of our computations. However, one needs to be careful to take into account different conventions used in defining the operators. In particular, to compare against \cite{Gehrmann:2023ksf} we need to divide their rules by a factor of $i^{N-1}/2$ (cf. e.g. their Eq.~(2.9)) while the rules in \cite{Klein:2009ig} need to be divided by $i^{N-1}$ (cf. their Eqs.~(2.86)-(2.87)).
\end{itemize}

\section{Derivation of the Feynman rule}
\label{sec:FeynRule}
In this section we provide the derivation of the operator vertices for an arbitrary number of derivatives. Without loss of generality we will focus on the operators $\Op_{k,0,N-k-1}$ for which the covariant derivatives only act on the $\psi$ field, cf.~Eq.~(\ref{eq:operators}). The Feynman rules for operators with covariant derivatives only acting on $\overline{\psi}$ can be obtained from these by exchanging $\delta_1 P_1$ and $\delta_2 P_2$ and setting $\delta_D\rightarrow -\delta_D$. Finally, operators with covariant derivatives acting on both $\psi$ and $\overline{\psi}$ can be written in terms of the other types using
\begin{equation}
\label{eq:partialAct}
    \Op_{p,q,r} = \Op_{p-1,q+1,r}+\Op_{p-1,q,r+1}.
\end{equation}
Before presenting the fully generic case, for which also an arbitrary number of gluons are attached to the operator vertex, we discuss the cases with zero, one and two gluons in order to clarify the procedure.

\subsection{LO Feynman rule}
At leading order, all covariant derivatives in the quark operator are simply replaced by partial ones,
\begin{equation}
\label{eq:LO1}
    \Op_{k,0,N-k-1}^{\text{LO}} = \partial^k[\overline{\psi}\Tilde{\Gamma}\partial^{N-k-1}\psi].
\end{equation}
To lighten the notation we introduced
\begin{equation}
    \Delta_{\mu}\partial^{\mu}\rightarrow \partial,\qquad \Delta_{\mu}D^{\mu}\rightarrow D,\qquad \Delta_{\mu}A^{\mu}\rightarrow A,\qquad \Delta_{\mu}\Gamma^{\mu}\rightarrow \Tilde{\Gamma}
\end{equation}
with $\Delta$ an arbitrary lightlike vector to select the leading-twist contributions. For the operators introduced below Eq.~(\ref{eq:generalOP}), $\Tilde{\Gamma}$ corresponds to $\slashed \Delta$, $\frac{1}{2}[\gamma_{\mu},\slashed\Delta]$ or $\gamma_5\slashed\Delta$ respectively. 
The derivatives in Eq.~(\ref{eq:LO1}) can be distributed using
\begin{equation}
\label{eq:distr}
    \partial^{k}(\phi_1\partial^{m}\phi_2) = \sum_{i=0}^{k}\binom{k}{i}(\partial^{k-i}\phi_1)(\partial^{m+i}\phi_2)
\end{equation}
which leads to
\begin{equation}
    \Op_{k,0,N-k-1}^{\text{LO}} = \sum_{i=0}^{k}\binom{k}{i}(\parder{k-i}\overline{\psi})\Tilde{\Gamma}(\parder{N-k-1+i}\psi).
\end{equation}
The Feynman rule for the LO operator vertex can now be found by Fourier transforming to momentum space. Using the rules in Eqs.~(\ref{eq:conventionsp1})-(\ref{eq:conventionsp2}) we find
\begin{figure}[H]
\centering
\includegraphics[width=0.8\textwidth,trim = 72 717 72 72]{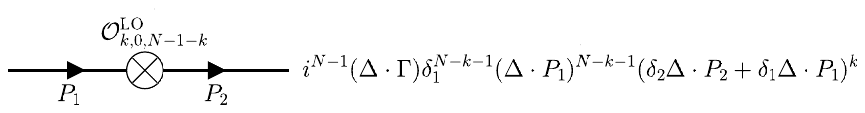}
\end{figure}
From this expression it is easy to see that the overall momentum scaling is $p^{N-1}$, as expected. The operators without total derivatives obey
\begin{equation}
    \Op_{0,0,N-1}^{\text{LO}} = i^{N-1}(\Delta\cdot\Gamma)\delta_{1}^{N-1}\deldotPPOW{1}{N-1}.
\end{equation}
Assuming all momenta to be incoming, $\delta_1=\delta_2=-1$, the latter agrees with Eq.~(A.2) of \cite{Gehrmann:2023ksf}. If instead we choose the physical momentum routing, $\delta_1=-1, \delta_2=+1$, we find find agreement with the LO rule in Fig.~(21) of \cite{Klein:2009ig}. Finally, it is easy to see that in both kinematic regimes the forward limit vanishes when $k>0$, as expected.

\subsection{NLO Feynman rule}
At next-to-leading order, one of the covariant derivatives is replaced by a gluon field. As such, one has to be careful to count all possible orderings of the derivatives and the gluon. To derive the Feynman rule, one can immediately start from $\Op_{k,0,N-k-1}$. Alternatively, using Eq.~(\ref{eq:partialAct}) we can write
\begin{equation}
\label{eq:kder}
    \Op_{k,0,N-k-1} = \sum_{i=0}^{k}\binom{k}{i}\Op_{0,k-i,N-k-1+i}
\end{equation}
and instead derive the Feynman rule for the operator on the right-hand side. Of course, the two methods have to lead to the same result. As illustration, we will now derive the NLO Feynman rule using the second approach.
As the gluon field can either come from the covariant derivatives acting on $\overline{\psi}$ or $\psi$ we have
\begin{align}
    \Op_{0,k,N-k-1}^{\text{NLO}} = -i\:g_s\delta_D\Bigg[&\sum_{i=0}^{k-1}\sum_{j=0}^{i}\binom{i}{j}(\parder{i-j}A)(\parder{k-1-i+j}\overline{\psi})\Tilde{\Gamma}(\parder{N-k-1}\psi)\nonumber\\&-(\parder{k}\overline{\psi})\Tilde{\Gamma}\sum_{i=0}^{N-k-2}\sum_{j=0}^{i}\binom{i}{j}(\parder{i-j}A)(\parder{N-k-2-i+j}\psi)\Bigg].
\end{align}
Next we apply Eq.~(\ref{eq:distr}) and transform to momentum space. Substituting into Eq.~(\ref{eq:kder}) and evaluating the sums then leads to
\begin{figure}[H]
\includegraphics[width=1\textwidth,trim = 72 702 72 72]{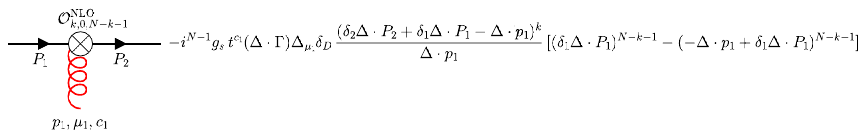}
\end{figure}
All the checks discussed at the end of Sec.~\ref{sec:setup} were explicitly performed. In particular, it is clear from the expression that the overall momentum scaling corresponds to $p^{N-2}$ as expected. Furthermore, for $k>0$ the operator vertex vanishes in the forward limit while for $k=0$, assuming all incoming momenta and choosing $\delta_D=+1$, our Feynman rule agrees with Eq.~(A.3) in \cite{Gehrmann:2023ksf}. For physical momentum routing our rule agrees with the NLO rule of Fig.~(21) in \cite{Klein:2009ig}. The LO and NLO Feynman rules (both for $\Gamma=\gamma_{\rho}$ and $\Gamma=\sigma_{\mu\rho}$) have been compared with a private code, kindly provided to one of the authors by J. Gracey, which implemented the rules for the $N=3$ operators with $k=0,1,2$. This code was then used in \cite{Gracey:2009da} for the computation and renormalization of the corresponding operator matrix elements. Complete agreement was found between our results and the expressions in that code. Finally, the LO and NLO Feynman rules for Wilson operators with an arbitrary number of total derivatives were presented in \cite{Kisselev:2005er,Kisselev:85.094022}, and we also agree with their results.

\subsection{NNLO Feynman rule}
Beyond NLO, multiple gluons are attached to the operator vertex. As such, one has to take into account all possible relative orderings. Of course in practice, one can simply compute the rule for one particular ordering of the gluon fields and then add all permutations of the color indices and momenta. This will be our approach here and in the following sections. We derive the rule starting directly with the operators with an arbitrary number of total derivatives
\begin{equation}
   \mathcal{O}_{k,0,N-k-1} = \parder{k}[\overline{\psi}\Tilde{\Gamma}D^{N-k-1}\psi].
\end{equation}
Replacing one of the covariant derivatives with a gluon field yields
\begin{equation}
\label{eq:NNLOa}
  {\mathcal{O}}_{k,0,N-k-1} \rightarrow -i\:g_s \delta_D \sum_{i=0}^{N-k-2}\parder{k}[\overline{\psi}\Tilde{\Gamma}D^{i}(A \:D^{N-k-2-i}\psi)].
\end{equation}
Next we select the second gluon. Note that we have a choice here: either it comes from the first $i$ covariant derivatives or it comes from the last $N-k-2-i$ ones. Of course, the Feynman rule obtained from these two options has to be exactly the same, assuming that the left-most gluon field is identified as $A_1^{\mu_1,c_1}(p_1)$ and the right-most one as $A_2^{\mu_2,c_2}(p_2)$. Choosing the second gluon to come from the first $i$ covariant derivatives for explicitness, we have
\begin{equation}
   \mathcal{O}_{k,0,N-k-1}^{\text{NNLO}} = -g_s^2 \delta_D^2 \sum_{i=0}^{N-k-2}\sum_{j=0}^{i-1}\parder{k}[\overline{\psi}\Tilde{\Gamma}\parder{j}(A_1 \parder{i-j-1}(A_2\parder{N-k-2-i}\psi))].
\end{equation}
After the iterative application of Eq.~(\ref{eq:distr}) this becomes
\begin{align}
\label{eq:NNLOb}
  \mathcal{O}_{k,0,N-k-1}^{\text{NNLO}} =  -g_s^2\delta_D^2\SUM{i=0}{N-k-2}\SUM{j=0}{i-1}\SUM{l=0}{k}\SUM{m=0}{j+l}&\SUM{n=0}{i-j+m-1}\binom{k}{l}\binom{j+l}{m}\binom{i-j+m-1}{n}\nonumber\\&\times(\parder{k-l}\overline{\psi})\Tilde{\Gamma}(\parder{N-k-2-i+n}\psi)(\parder{j+l-m}A_1)(\parder{i-j+m-n-1}A_2).
\end{align}
Transforming to momentum space and evaluating the sums we then find the following Feynman rule for the NNLO operator vertex
\begin{figure}[H]
\includegraphics[width=1\textwidth,trim = 72 683 72 72]{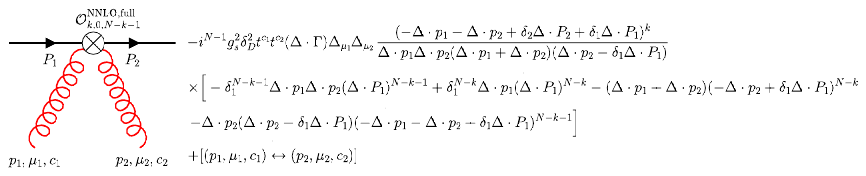}
\end{figure}
If we had chosen the second gluon in Eq.~(\ref{eq:NNLOa}) to come from the last $N-k-2-i$ covariant derivatives instead, Eq.~(\ref{eq:NNLOb}) would be replaced by
\begin{align}
\label{eq:NNLOc}
 \mathcal{O}_{k,0,N-k-1}^{\text{NNLO}} =  -g_s^2\delta_D^2\SUM{i=0}{N-k-2}\SUM{j=0}{N-k-3-i}\SUM{l=0}{k}\SUM{m=0}{i+l}&\SUM{n=0}{j+m}\binom{k}{l}\binom{i+l}{m}\binom{j+m}{n}\delta_1^{k-l}\delta_2^{N-k-3-i-j+n}\nonumber\\&\times(\parder{k-l}\overline{\psi})\Tilde{\Gamma}(\parder{N-k-3-i-j+n}\psi)(\parder{i+l-m}A_1)(\parder{j+m-n}A_2).
\end{align}
It is straightforward to check that the corresponding Feynman rule is equivalent to the one presented above. Furthermore, we have explicitly implemented all the checks listed at the end of Sec.~\ref{sec:setup}. In particular the operator vertex vanishes in the forward limit if $k>0$ while for $k=0$ it agrees with Eq.~(A.4) in \cite{Gehrmann:2023ksf} for all momenta incoming and with the NNLO rule of Fig.~(21) in \cite{Klein:2009ig} for physical momentum routing. By setting $\Gamma=\gamma_5\gamma_{\rho}$ our rules for $k=0$ also agree with those for the polarized operators, see e.g.~\cite{Mertig:1995ny}. 

\subsection{N$^K$LO Feynman rule}
We now generalize the above results to operator vertices with an arbitrary number of total derivatives $k$ and an arbitrary number of gluons $K$. As mentioned before, such vertices become relevant for $K$-loop computations. The first step is to determine all possible orderings of the $K$ gluons and the $N-k-K-1$ derivatives. We find
\begin{align}
    \Op_{k,0,N-k-1}^{\text{N}^K\text{LO}} = (-i\:g_s\delta_D)^{K}\sum_{i_1=0}^{N-k-2}\prod_{j=1}^{K-1}\sum_{i_{j+1}=0}^{i_{j}-1}\parder{k}[\overline{\psi}\Tilde{\Gamma}\parder{i_{K}}(A_{j}\parder{i_{j}-i_{j+1}-1}(A_{K}\parder{N-k-i_{1}-2}\psi))].
\end{align}
Next we need to distribute all total derivatives. This can be done by iteratively applying Eq.~(\ref{eq:distr}), leading to
\begin{align}
\label{eq:genericOP}
    \Op_{k,0,N-k-1}^{\text{N}^K\text{LO}} = (-i\:g_s\delta_D)^{K}&\sum_{l=0}^{k}\binom{k}{l}(\parder{k-l}\overline{\psi})\Tilde{\Gamma}\prod_{j=0}^{K-1}\sum_{i_{j+1}=0}^{i_{j}-1}\sum_{m_{j+1}=0}^{i_{j}-i_{j+1}+m_{j}-1}\binom{i_{j}-i_{j+1}+m_{j}-1}{m_{j+1}}\nonumber\\&\times(\parder{i_{j}-i_{j+1}+m_{j}-m_{j+1}-1}A_{j+1})(\parder{N-k-i_{1}+m_{K}-2}\psi).
\end{align}
Here we introduced the auxiliary symbols $i_0$ and $m_0$ defined as
\begin{equation}
    i_0-1 = N-k-2
\end{equation}
and
\begin{equation}
\label{eq:m0}
    i_0-i_1+m_0-1=i_K+l.
\end{equation}
Note that in practice, the product over the sums is to be interpreted as
\begin{equation}
    \prod_{j}\sum_{i_{j+1}}\sum_{m_{j+1}} \sim \left(\prod_{j}\sum_{i_{j+1}} \right)\left(\prod_{r}\sum_{m_{r+1}} \right),
\end{equation}
i.e. the product of $i$-sums needs to be expanded before the one over $m$-sums. The reason for this is that the upper index of the first $m$-sum actually involves $i_K$, cf. Eq.~(\ref{eq:m0}). From Eq.~(\ref{eq:genericOP}), it now follows that the N$^K$LO Feynman rule is given by
\begin{figure}[H]
\centering
\includegraphics[width=1\textwidth,trim = 72 643 72 72]{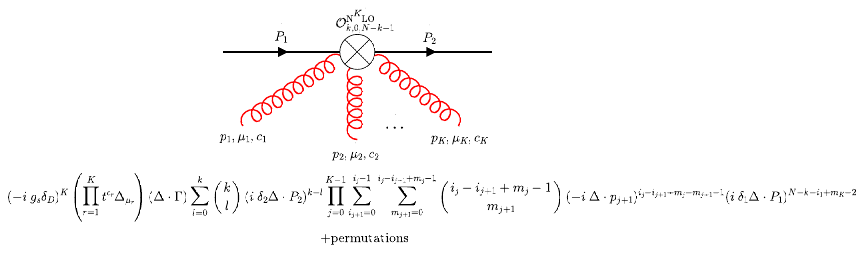}
\end{figure}
\noindent Here ``+ permutations'' denotes the fact that all permutations of the gluons have to be added to obtain the full rule. The generic rule reproduces the LO, NLO and NNLO results presented above for $K=0$, $K=1$ and $K=2$ respectively. Furthermore, for the operators without total derivatives, $k=0$, we find agreement with the rules available in the literature up to N$^3$LO, cf. e.g. Eqs.~(A.2)-(A.5) in \cite{Gehrmann:2023ksf} for all momenta coming into the vertex and Fig.~(21) in \cite{Klein:2009ig} for physical momentum routing. We also agree with the N$^4$LO vertex presented in \cite{Moch:2017uml}, cf. their Eq.~(A.6) \footnote{Note however that the rules in \cite{Moch:2017uml} are missing some powers of $i$.}. Our results are also consistent with \cite{Mikhailov:2020tta}, in which the corresponding Feynman rule is presented in $x$-space. Finally, for fixed values of $N$, $K$ and $k>0$, it is straightforward to check that the forward limit of the operator vertex vanishes, as expected.

\subsection{Extension to gluon operators}
\label{sec:FeynRuleGluon}
The general form of the all-order Feynman rule for the gauge-invariant quark operator, and its corresponding operator form in Eq.~(\ref{eq:genericOP}), can be used as a starting point in the derivation of the Feynman rules of other types of operators. As illustration, we present in this section the Feynman rule for the gauge-invariant leading-twist gluon operator
\begin{equation}
    \Op_{k,0,N-k-2}^G = \mathcal{S}\partial_{\mu_1}\dots\partial_{\mu_k}[F_{\nu\rho_1} D_{\rho_{2}}\dots D_{\rho_{N-k-3}} F_{\rho_{N-k-2}}^{}{}^{\nu}].
\end{equation}
We use a similar convention for the covariant derivative in the adjoint representation as before, i.e. 
\begin{equation}
    D_{\mu}F_{\rho\sigma}^{a} = \partial_{\mu}F_{\rho\sigma}^{a}-\delta_{D}(i g_s T^{a}_{bc}A_{\mu}^{b}F_{\rho\sigma}^{c}).
\end{equation}
Using $T^{a}_{bc}=i f^{abc}$, with $f^{abc}$ the standard structure constants of $SU(N_c)$, this becomes
\begin{equation}
\label{eq:covDGlu}
    D_{\mu}F_{\rho\sigma}^{a} = \partial_{\mu}F_{\rho\sigma}^{a}+\delta_{D} (g_s f^{abc}A_{\mu}^{b}F_{\rho\sigma}^{c}).
\end{equation}
With these conventions the gluon field strength takes the following form
\begin{equation}
\label{eq:fieldstr}
    F_{\mu\nu}^{a} = \partial_{\mu}A_{\nu}^{a}-\partial_{\nu}A_{\mu}^{a}+\delta_D (g_s f^{abc}A_{\mu}^{b}A_{\nu}^{c}).
\end{equation}
As we did above, we implement the leading-twist approximation by contracting the operator with $N$ copies of a lightlike vector $\Delta$
\begin{equation}
\label{eq:gluonOP}
     \Op_{k,0,N-k-2}^G = \partial^{k}[F_{\nu\Delta} D^{N-k-2} F_{\Delta}^{}{}^{\nu}]
\end{equation}
in which we defined
\begin{equation}
    F_{\mu\Delta} \equiv F_{\mu\lambda}\Delta^{\lambda}.
\end{equation}
We mimic the notation introduced above and write the operator vertex associated to Eq.~(\ref{eq:gluonOP}) as
\begin{figure}[H]
\centering
\includegraphics[width=0.5\textwidth,trim = 72 502 72 72]{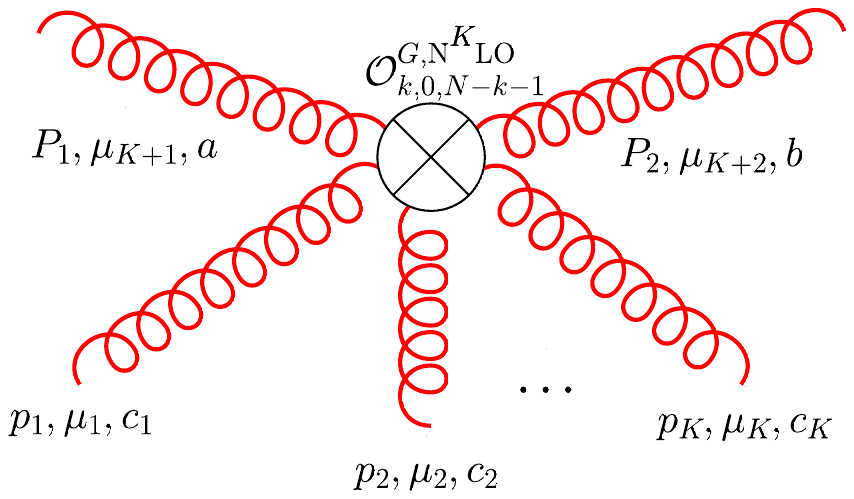}
\end{figure}
\noindent i.e. two of the gluons are denoted with $A_{\mu_{K+1}}^{a}(P_1)$ and $A_{\mu_{K+2}}^{b}(P_2)$ while each extra gluon at higher orders is denoted by $A_{\mu_{i}}^{c_i}(p_i)$ with $i\in[1,K]$. To derive the corresponding Feynman rule, we now follow the same reasoning as for the derivation of the quark operator rule. In fact, we can recycle the expression presented in Eq.~(\ref{eq:genericOP}) for the gluon operator with some simple replacements
\begin{itemize}
    \item $N\rightarrow N-1$, to take into account that the gluon operator has one less covariant derivative than the quark one,
    \item $\Tilde{\Gamma}\rightarrow 1$,
    \item $(-i\:g_s\Delta_D)^{K} \rightarrow (g_s\Delta_D)^{K}$, cf.~Eq.~(\ref{eq:covDGlu}),
    \item $\parder{k-l}\overline{\psi}\rightarrow\parder{k-l}F_{\nu\Delta}$,
    \item $\parder{N-k-i_1+m_K-2}\psi\rightarrow\parder{N-k-i_1+m_K-3}F_{\Delta}^{}{}^{\nu}$.
\end{itemize}
This implies that the $K$-th order gluon operator is written as
\begin{align}
\label{eq:genericgluonOP}
    \Op_{k,0,N-k-2}^{G,\text{N}^K\text{LO}} = (g_s\delta_D)^{K}&\sum_{l=0}^{k}\binom{k}{l}(\parder{k-l}F_{\nu\Delta})\prod_{j=0}^{K-1}\sum_{i_{j+1}=0}^{i_{j}-1}\sum_{m_{j+1}=0}^{i_{j}-i_{j+1}+m_{j}-1}\binom{i_{j}-i_{j+1}+m_{j}-1}{m_{j+1}}\nonumber\\&\times(\parder{i_{j}-i_{j+1}+m_{j}-m_{j+1}-1}A_{j+1})(\parder{N-k-i_{1}+m_{K}-3}F_{\Delta}^{}{}^{\nu})
\end{align}
where, as in Eq.~(\ref{eq:genericOP}), we have stripped off the overall color factor $\mathcal{C}^{a b c_1 \dots c_K}$. Because of the structure of the gluon operator, this color factor can be written as
\begin{equation}
\label{eq:color}
    \mathcal{C}^{a b c_1 \dots c_K} = \prod_{j=0}^{K-1}f^{x_{j}c_{j+1}x_{j+1}}
\end{equation}
with $x_0 \equiv a$ and $x_K \equiv b$.\newline

Note however that, since the field strength depends on the coupling, cf.~Eq.~(\ref{eq:fieldstr}), the terms in Eq.~(\ref{eq:genericgluonOP}) have to be reorganized to select those that are $O(g_s^{K})$. To do so, we need to take into account three contributions
\begin{itemize}
    \item double Abelian (Ab-Ab): the product of the Abelian parts of the field strengths, substituted in the $K$-th order vertex,
    \item Abelian-non-Abelian (Ab-nAb): the product of an Abelian term with a  non-Abelian one, substituted in the $(K-1)$-st order vertex and
    \item double-non-Abelian (nAb-nAb): the product of the two non-Abelian parts of the field strengths, substituted in the $(K-2)$-nd order vertex.
\end{itemize}
Accordingly we write
\begin{equation}
\label{eq:gluonDecomp}
    \Op_{k,0,N-k-2}^{G,\text{N}^K\text{LO}} = \Op_{k,0,N-k-2}^{\text{Ab-Ab},\text{N}^K\text{LO}}+\Op_{k,0,N-k-2}^{\text{Ab-nAb},\text{N}^K\text{LO}}+\Op_{k,0,N-k-2}^{\text{nAb-nAb},\text{N}^K\text{LO}}
\end{equation}
in which each term is now $O(g_s^K)$.
As illustration, let us consider the first contribution which corresponds to replacing the field strengths by their Abelian parts. Because of the structure of the operator in Eq.~(\ref{eq:genericgluonOP}), we concentrate on a product of field strengths of the form $\parder{N}(F_{\nu\Delta}^{a})\parder{M}(F_{\Delta\nu}^{b})$ for arbitrary values of $N$ and $M$. We have
\begin{equation}
    \parder{N}(F_{\nu\Delta}^{a})\parder{M}(F_{\Delta\nu}^{b}) = (\partial_{\nu}\parder{N}A^{a}-\parder{N+1}A_{\nu}^{a})(\parder{M+1}A_{\nu}^{b}-\partial_{\nu}\parder{M}A^{b}) \equiv \Op_{\mu\nu}^{ab}(N,M)
\end{equation}
such that the corresponding Feynman rule becomes
\begin{figure}[H]
\includegraphics[width=1\textwidth, trim = 72 738 72 72]{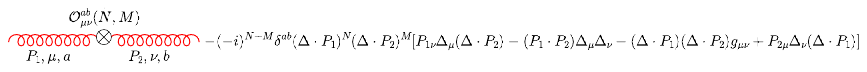}
\end{figure}
At higher orders more gluons are attached to this vertex coming from the expansion of the covariant derivatives in the operator. Following the notation introduced above, this gives an additional factor that, symbolically, is of the form $\prod_{j=1}^{K}\Delta_{\mu_j}(\Delta\cdot p_j)^{f(j)}$. Finally we need to substitute everything into the $K$-th order vertex, cf.~Eq.~(\ref{eq:genericgluonOP}), replacing $N$, $M$ and $f(j)$ by the appropriate functions. Furthermore the $\delta^{ab}$ should be replaced by $\mathcal{C}^{ab c_1 \dots c_K}$ for the $(K+2)$-gluon vertex, cf.~Eq.~(\ref{eq:color}). We find
\begin{align}
\label{eq:gluonAbelian}
    &\Op_{k,0,N-k-2}^{\text{Ab-Ab},\text{N}^K\text{LO}} = (g_s\delta_D)^{K}\left(\prod_{r=1}^{K}\Delta_{\mu_r}\right)\mathcal{C}^{a b c_1\dots c_K}\mathcal{L}_{\mu_{K+1}\mu_{K+2}}^{(1)}\sum_{l=0}^{k}\binom{k}{l}(-i\:\Delta\cdot P_1)^{k-l}\nonumber\\&\times\prod_{j=0}^{K-1}\sum_{i_{j+1}=0}^{i_{j}-1}\sum_{m_{j+1}=0}^{i_{j}-i_{j+1}+m_{j}-1}\binom{i_{j}-i_{j+1}+m_{j}-1}{m_{j+1}}(-i\:\Delta\cdot p_{j+1})^{i_{j}-i_{j+1}+m_{j}-m_{j+1}-1}\nonumber\\&\times(-i\:\Delta\cdot P_2)^{N-k-i_{1}+m_{K}-3}  + \text{\: permutations}
\end{align}
with
\begin{align}
    \mathcal{L}_{\mu_{K+1}\mu_{K+2}}^{(1)} =& -P_{1\mu_{K+2}}\Delta_{\mu_{K+1}}(\Delta\cdot P_2)+(P_1\cdot P_2)\Delta_{\mu_{K+1}}\Delta_{\mu_{K+2}}+(\Delta\cdot P_1)(\Delta\cdot P_2)g_{\mu_{K+1}\mu_{K+2}}\nonumber\\&-P_{2\mu_{K+1}}\Delta_{\mu_{K+2}}(\Delta\cdot P_1)
\end{align}
and
\begin{align}
    &i_0-1 = N-k-3, \\
    &i_0-i_1+m_0-1=i_K+l.
\end{align}
The `+ permutations' in this rule denotes the fact that all permutations of the gluon fields have to be added. As the number of gluons in the $K$-th order vertex is $(K+2)$, the corresponding Feynman rule needs to have all $(K+2)!$ permutations~\footnote{Note however that not all of these terms are independent as e.g. $f^{abc}=-f^{acb}$.}. The treatment of the remaining two terms in Eq.~(\ref{eq:gluonDecomp}) works in a similar way and we find
\begin{align}
\label{eq:gluonNAbelian}
    &\Op_{k,0,N-k-2}^{\text{Ab-nAb},\text{N}^K\text{LO}} = (g_s\delta_D)^{K}\left(\prod_{r=1}^{K}\Delta_{\mu_r}\right)\mathcal{C}^{a b c_1\dots c_K}\mathcal{L}_{\mu_{K+1}\mu_{K+2}}^{(2)}\sum_{l=0}^{k}\binom{k}{l}(-i\: \Delta\cdot P_1-i\: \Delta\cdot p_1)^{k-l}\nonumber\\&\times\prod_{j=0}^{K-2}\sum_{i_{j+1}=0}^{i_{j}-1}\sum_{m_{j+1}=0}^{i_{j}-i_{j+1}+m_{j}-1}\binom{i_{j}-i_{j+1}+m_{j}-1}{m_{j+1}}(-i\:\Delta\cdot p_{j+2})^{i_{j}-i_{j+1}+m_{j}-m_{j+1}-1}\nonumber\\&\times(-i\:\Delta\cdot P_2)^{N-k-i_{1}+m_{K-1}-3}  + \text{\: permutations}
\end{align}
with
\begin{align}
    &\mathcal{L}_{\mu_{K+1}\mu_{K+2}}^{(2)} = -2i\left[g_{\mu_{K+1}\mu_{K+2}}(\Delta\cdot P_2)-P_{2\mu_{K+1}}\Delta_{\mu_{K+2}}\right], \\
    &i_0-1 = N-k-3, \\
    &i_0-i_1+m_0-1=i_{K-1}+l
\end{align}
and
\begin{align}
\label{eq:gluonDAbelian}
    &\Op_{k,0,N-k-2}^{\text{nAb-nAb},\text{N}^K\text{LO}} = (g_s\delta_D)^{K}\left(\prod_{r=1}^{K}\Delta_{\mu_r}\right)\mathcal{C}^{a b c_1\dots c_K}\mathcal{L}_{\mu_{K+1}\mu_{K+2}}^{(3)}\sum_{l=0}^{k}\binom{k}{l}(-i\:\Delta\cdot P_1-i\:\Delta\cdot p_1)^{k-l}\nonumber\\&\times\prod_{j=0}^{K-3}\sum_{i_{j+1}=0}^{i_{j}-1}\sum_{m_{j+1}=0}^{i_{j}-i_{j+1}+m_{j}-1}\binom{i_{j}-i_{j+1}+m_{j}-1}{m_{j+1}}(-i\:\Delta\cdot p_{j+2})^{i_{j}-i_{j+1}+m_{j}-m_{j+1}-1}\nonumber\\&\times(-i\:\Delta\cdot P_2-i\:\Delta\cdot p_{K})^{N-k-i_{1}+m_{K-2}-3}  + \text{\: permutations}
\end{align}
with
\begin{align}
    &\mathcal{L}_{\mu_{K+1}\mu_{K+2}}^{(3)} = g_{\mu_{K+1}\mu_{K+2}}, \\
    &i_0-1 = N-k-3, \\
    &i_0-i_1+m_0-1=i_{K-2}+l.
\end{align}
Hence the full $K$-th order gluon Feynman rule is now given by Eq.~(\ref{eq:gluonDecomp}) with the three terms in the decomposition given by Eqs.~(\ref{eq:gluonAbelian}), (\ref{eq:gluonNAbelian}) and (\ref{eq:gluonDAbelian}). We have performed the following checks on our result:
\begin{itemize}
    \item from the functional form of the operator, the overall momentum scaling of the Feynman rule is expected to be $p^{N-K}$, which our rule indeed obeys;
    \item the expression vanishes in the forward limit when $k>0$;
    \item we have checked that the results up to N$^3$LO match the literature, explicitly Eqs.~(A.6)-(A.9) in \cite{Gehrmann:2023ksf}~\footnote{To take into account different conventions in the operator definitions, the rules in \cite{Gehrmann:2023ksf} should be divided by $(-i)^{N-2}/2$ when comparing to our expressions, cf. their Eq.~(2.9).}.
\end{itemize}


Finally, note that the above discussion is easily generalized to the polarized gluon operator, defined as
\begin{equation}
    \Tilde{\Op}_{k,0,N-k-2}^G = \mathcal{S}\partial_{\mu_1}\dots\partial_{\mu_k}[\Tilde{F}_{\nu\rho_1} D_{\rho_{2}}\dots D_{\rho_{N-k-3}} F_{\rho_{N-k-2}}^{}{}^{\nu}]
\end{equation}
with
\begin{equation}
    \Tilde{F}_{\mu\nu} = \frac{1}{2}\eps_{\mu\nu\rho\sigma}F^{\rho\sigma}
\end{equation}
the gluon dual field strength and $\eps_{\mu\nu\rho\sigma}$ the standard completely anti-symmetric Levi-Civita symbol. Performing the same steps as for the unpolarized operator we find that the Feynman rule is of the form
\begin{equation}
    \Tilde{\Op}_{k,0,N-k-2}^{G,\text{N}^K\text{LO}} = \Tilde{\Op}_{k,0,N-k-2}^{\text{Ab-Ab},\text{N}^K\text{LO}}+\Tilde{\Op}_{k,0,N-k-2}^{\text{Ab-nAb},\text{N}^K\text{LO}}+\Tilde{\Op}_{k,0,N-k-2}^{\text{nAb-nAb},\text{N}^K\text{LO}}
\end{equation}
with the three contributions defined as follows
\begin{itemize}
    \item[(1)] \begin{align}
\label{eq:gluonAbelianPOL}
    &\Tilde{\Op}_{k,0,N-k-2}^{\text{Ab-Ab},\text{N}^K\text{LO}} = (g_s\delta_D)^{K}\left(\prod_{r=1}^{K}\Delta_{\mu_r}\right)\mathcal{C}^{a b c_1\dots c_K}\Tilde{\mathcal{L}}_{\mu_{K+1}\mu_{K+2}}^{(1)}\sum_{l=0}^{k}\binom{k}{l}(-i\:\Delta\cdot P_1)^{k-l}\nonumber\\&\times\prod_{j=0}^{K-1}\sum_{i_{j+1}=0}^{i_{j}-1}\sum_{m_{j+1}=0}^{i_{j}-i_{j+1}+m_{j}-1}\binom{i_{j}-i_{j+1}+m_{j}-1}{m_{j+1}}(-i\:\Delta\cdot p_{j+1})^{i_{j}-i_{j+1}+m_{j}-m_{j+1}-1}\nonumber\\&\times(-i\:\Delta\cdot P_2)^{N-k-i_{1}+m_{K}-3}  + \text{\: permutations}
\end{align}
with
\begin{align}
    &\Tilde{\mathcal{L}}_{\mu_{K+1}\mu_{K+2}}^{(1)} = (\Delta\cdot P_2)\eps_{\mu_{K+2}\:\Delta\: P_1\:\mu_{K+1}}-\Delta_{\mu_{K+2}}\eps_{P_2\:\Delta\: P_1\:\mu_{K+1}}, \\
    &i_0-1 = N-k-3, \\
    &i_0-i_1+m_0-1=i_K+l,
\end{align}

\item[(2)] \begin{align}
\label{eq:gluonNAbelianPOL}
    &\Tilde{\Op}_{k,0,N-k-2}^{\text{Ab-nAb},\text{N}^K\text{LO}} = (g_s\delta_D)^{K}\left(\prod_{r=1}^{K-1}\Delta_{\mu_r}\right)\mathcal{C}^{a b c_1\dots c_K}\Tilde{\mathcal{L}}_{\mu_K\mu_{K+1}\mu_{K+2}}^{(2a)}\sum_{l=0}^{k}\binom{k}{l}(-i\: \Delta\cdot P_2-i\: \Delta\cdot p_K)^{k-l}\nonumber\\&\times\prod_{j=0}^{K-2}\sum_{i_{j+1}=0}^{i_{j}-1}\sum_{m_{j+1}=0}^{i_{j}-i_{j+1}+m_{j}-1}\binom{i_{j}-i_{j+1}+m_{j}-1}{m_{j+1}}(-i\:\Delta\cdot p_{K-j-1})^{i_{j}-i_{j+1}+m_{j}-m_{j+1}-1}\nonumber\\&\times(-i\:\Delta\cdot P_1)^{N-k-i_{1}+m_{K-1}-3}  + (g_s\delta_D)^{K}\left(\prod_{r=1}^{K}\Delta_{\mu_r}\right)\mathcal{C}^{a b c_1\dots c_K}\Tilde{\mathcal{L}}_{\mu_{K+1}\mu_{K+2}}^{(2b)}\sum_{l=0}^{k}\binom{k}{l}(-i\: \Delta\cdot P_2)^{k-l}\nonumber\\&\times\prod_{j=0}^{K-2}\sum_{i_{j+1}=0}^{i_{j}-1}\sum_{m_{j+1}=0}^{i_{j}-i_{j+1}+m_{j}-1}\binom{i_{j}-i_{j+1}+m_{j}-1}{m_{j+1}}(-i\:\Delta\cdot p_{K-j})^{i_{j}-i_{j+1}+m_{j}-m_{j+1}-1}\nonumber\\&\times(-i\:\Delta\cdot P_1-i\:p_1)^{N-k-i_{1}+m_{K-1}-3} + \text{\: permutations}
\end{align}
with
\begin{align}
    &\Tilde{\mathcal{L}}_{\mu_{K}\mu_{K+1}\mu_{K+2}}^{(2a)} = (-1)^{K}\:\frac{i}{2}\left(\eps_{\mu_{K+1}\:\Delta\:\mu_{K+2}\:\mu_K}(\Delta\cdot P_1)-\eps_{P_1\:\Delta\:\mu_{K+2}\:\mu_K}\Delta_{\mu_{K+1}}\right), \\
    &\Tilde{\mathcal{L}}_{\mu_{K+1}\mu_{K+2}}^{(2b)} = (-1)^{K}\:i\:\eps_{\mu_{K+1}\:\Delta\:P_2\:\mu_{K+2}}, \\
    &i_0-1 = N-k-3, \\
    &i_0-i_1+m_0-1=i_{K-1}+l
\end{align}

\item[(3)] \begin{align}
\label{eq:gluonDAbelianPOL}
    &\Tilde{\Op}_{k,0,N-k-2}^{\text{nAb-nAb},\text{N}^K\text{LO}} = (g_s\delta_D)^{K}\left(\prod_{r=2}^{K}\Delta_{\mu_r}\right)\mathcal{C}^{a b c_1\dots c_K}\Tilde{\mathcal{L}}_{\mu_1\mu_{K+1}\mu_{K+2}}^{(3)}\sum_{l=0}^{k}\binom{k}{l}(-i\:\Delta\cdot P_1-i\:\Delta\cdot p_1)^{k-l}\nonumber\\&\times\prod_{j=0}^{K-3}\sum_{i_{j+1}=0}^{i_{j}-1}\sum_{m_{j+1}=0}^{i_{j}-i_{j+1}+m_{j}-1}\binom{i_{j}-i_{j+1}+m_{j}-1}{m_{j+1}}(-i\:\Delta\cdot p_{j+2})^{i_{j}-i_{j+1}+m_{j}-m_{j+1}-1}\nonumber\\&\times(-i\:\Delta\cdot P_2-i\:\Delta\cdot p_{K})^{N-k-i_{1}+m_{K-2}-3} + \text{\: permutations}
\end{align}
with
\begin{align}
    &\Tilde{\mathcal{L}}_{\mu_1\mu_{K+1}\mu_{K+2}}^{(3)} = -\frac{(-1)^{K}}{2}\eps_{\mu_{K+2}\:\Delta\:\mu_{K+1}\:\mu_1}, \\
    &i_0-1 = N-k-3, \\
    &i_0-i_1+m_0-1=i_{K-2}+l.
\end{align}
\end{itemize}
We have explicitly checked that, for $k=0$, our LO and NLO results agree with the rules presented in \cite{Mertig:1995ny} \footnote{Because of differences in the operator definitions, the rules in \cite{Mertig:1995ny} should be divided by $i^{N+1}$ when comparing to our expressions.}. Note however that the comparison is not direct, as one needs to use the Schouten identity which reads
\begin{equation}
\label{eq:Schouten}
V^{\tau}\eps^{\mu\nu\rho\sigma}+V^{\mu}\eps^{\nu\rho\sigma\tau}+V^{\nu}\eps^{\rho\sigma\tau\mu}+V^{\rho}\eps^{\sigma\tau\mu\nu}+V^{\sigma}\eps^{\tau\mu\nu\rho} = 0
\end{equation}
for some arbitrary vector $V$. The NNLO result in \cite{Mertig:1995ny} is incorrect, as pointed out already in \cite{Behring:2019tus}. In the latter paper the authors present a corrected version, cf. their Eq.~(144). However, it turns out that also this expression is incorrect. One way to see this is that it is not permutation invariant. Interestingly, it seems that one should combine different parts of the expressions in \cite{Mertig:1995ny} and \cite{Behring:2019tus}. Both agree for the double-non-Abelian part, corresponding to the terms without any sums. Then however one should take the double-Abelian part, corresponding to the terms with the double sums, from \cite{Mertig:1995ny} while from \cite{Behring:2019tus} we need to take the Abelian-non-Abelian piece, i.e. the terms containing the single sums. Our rule then correctly reproduces the resulting expression. The expression for the five-gluon vertex is also known, cf.~Eqs.~(A.1)-(A.3) in \cite{Blumlein:2021ryt}. Dividing out an overall factor of $-i/(2\:i^{N-2})$, coming from differing operator definitions, we agree with their rule up to a factor of $1/2$ in the double-non-Abelian part (i.e. the terms with only single sums). Finally, we explicitly checked that the operator vertex vanishes in the forward limit when $k>0$ and that it vanishes when $N$ is even \footnote{To show this, one needs to take into account all the gluon permutations and use the Schouten identity, cf.~Eq.(\ref{eq:Schouten}).}.


\section{Implementation of the N$^K$LO Feynman rules}
\label{sec:implementation}
The expressions above for the quark and gluon Feynman rules are written in a form appropriate for implementation in computer algebra systems. As illustration, we provide implementations both in {\tt Mathematica} and in \textit{{\tt FORM}}, which are available at \url{https://github.com/vtsam/NKLO}. Below we briefly summarize how these codes can be used.\newline

\textbf{\underline{{\tt Mathematica}}}
The NKLO.wl file contain two functions, called {\tt NKLO} and {\tt NKLOg}, which take the arguments {\tt N}, {\tt k} and {\tt K} (in that order). The {\tt NKLO} function encodes the Feynman rule for the quark operators and accepts the following options:
\begin{itemize}
    \item {\texttt{kinematics}}: choose the momentum routing, which corresponds to setting $\delta_1$ and $\delta_2$ above to specific values. The default value is "{\tt generic}", which keeps $\delta_1$ and $\delta_2$ symbolic. Other allowed values are "{\tt incoming}", which sets $\delta_1=\delta_2=-1$ and "{\tt physical}" for which $\delta_1=-1, \delta_2=+1$.
    \item {\tt covD}: specifies the convention used for the covariant derivative, which corresponds to setting a value for $\delta_D$ above. The default option is "{\tt generic}", which keeps $\delta_D$ symbolic. Alternatively one can set "{\tt covD}" $\rightarrow$ {\tt 1} or "{\tt covD}" $\rightarrow$ {\tt -1}.
    \item {\tt perms}: include all permutations of the gluons. When set to {\tt True}, all $K!$ permutations are included. When set to {\tt False}, the rule only keeps one particular ordering, corresponding to $t^{c_1}\dots t^{c_K}$. The default value is {\tt True}.
\end{itemize}
Furthermore, the output of the {\tt NKLO} function is such that the momenta are written in the same way as in the text, i.e. the quark momenta are {\tt P[1]} and {\tt P[2]} while the momenta of the additional gluons are denoted by {\tt p[i]}. The scalar product of the lightlike $\Delta$ with some momentum $r$ is denoted as {\tt Delta[r]}. Note that the input parameters need not all be numeric. In particular, one can produce all-$N$ expressions by keeping the first argument of the {\tt NKLO} function symbolic. For example, the function call
\begin{framed}
    {\small{\tt \textbf{In[1]:= NKLO[N,0,1,"{\color{gray}perms}" $\rightarrow$ False,"{\color{gray}kinematics}" $\rightarrow$ "{\color{gray}incoming}"]}}}
\end{framed}
\noindent will reproduce the NLO Feynman rule. Such symbolic expression are useful when one wants to perform the all-$N$ reductions generically. This can for example be achieved by resumming the operator insertions into linear propagators, see e.g.~\cite{Ablinger:2012qm,Ablinger:2014yaa}. The code can also produce the rules for symbolic $k$ and symbolic ($N$, $k$). The number of gluons however always needs to be specified explicitly (i.e. $K\geq 0$). For example, the Feynman rule for the NNLO vertex can be generated with the function call~\footnote{For the generation of symbolic Feynman rules, we suggest to use the option "{\tt perms}" $\rightarrow$ {\tt False} to speed up the computation.}
\begin{framed}
    {\small{\tt \textbf{In[1]:= NKLO[N,k,2,"{\color{gray}perms}" $\rightarrow$ False]}}}
\end{framed}
The second provided function, {\tt NKLOg}, encodes the Feynman rule for the gluon operators. Here the following options are available:
\begin{itemize}
    \item {\tt polarization}: choose whether to generate the rules for the polarized or unpolarized gluon operators. The default value is {\tt False}.
    \item {\tt Abelian}: choose whether to work in an Abelian or a non-Abelian theory. When set to {\tt True}, only the first term in Eq.~(\ref{eq:gluonDecomp}), which corresponds to taking into account only the Abelian part of the field strength, is kept \footnote{When setting this option to {\tt True}, one needs to be careful with the interpretation of the prefactors, which are written for QCD. This responsibility is put on the user.}. The default value is {\tt False}, meaning the generated rule is valid for a non-Abelian model like QCD.
    \item {\tt covD}: specifies the convention used for the covariant derivative, which corresponds to setting a value for $\delta_D$ in Eq.~(\ref{eq:covDGlu}) above. The default option is {\tt generic}, which keeps $\delta_D$ symbolic. Alternatively one can set "{\tt covD}" $\rightarrow$ {\tt 1} or "{\tt covD}" $\rightarrow$ {\tt -1}.
    \item {\tt perms}: include all permutations of the gluons. When set to {\tt True}, all $(K+2)!$ permutations are included. Furthermore, the anti-symmetry properties of the structure constants are implemented to obtain a minimal set of color structures in the output. When set to {\tt False}, the rule only keeps one particular ordering corresponding to $\mathcal{C}^{a b c_1 \dots c_K}$, cf.~Eq.~(\ref{eq:color}). The default value is {\tt True}.
\end{itemize}

\textbf{\underline{{\tt FORM}}} We also provide a \textit{{\tt FORM}} code to generate the operator Feynman rules. This can be used by running the {\tt NKLO.sh} script. When run, the user will be prompted to input the operator type, operator spin, the number of total derivatives and the number of additional gluons. Contrary to the {\tt Mathematica} implementation, the last three values need to be numeric. Valid input options for the operator type are q for the quark operators, g for the unpolarized gluon operators and gp for the polarized ones. Furthermore, the user is asked whether or not to include permutations. When the latter is set to 0, the part of the Feynman rule corresponding to the color-ordering $t^{c_1}\dots t^{c_K}$ is generated for the quark operators and $f^{a c_1 x_1}f^{x_1 c_1 x_2}\dots f^{x_{K-1}c_K b}$ for the gluon operators. All other permutations are omitted. If the value for the permutations is different from 0, all permutations of the gluonic fields are generated. The rule is written to the output file output\_...op\_N...\_k...\_K... .h (e.g.~the rule for the $N=6$ quark operator with $k=1$ and $K=2$ is written to the file output\_qop\_N6\_k1\_K2.h).

\section{Conclusions and outlook}
\label{sec:conclusion}
In this article, we derived the Feynman rules for leading-twist gauge invariant operators with an arbitrary number of total derivatives and an arbitrary number of gluons. The knowledge of such Feynman rules is important for the study of hard QCD scattering processes, as the scale dependence of parton distributions is determined by the anomalous dimensions of the operators that define these distributions. The anomalous dimensions can be computed perturbatively by renormalizing the partonic matrix elements of the operators, which requires the Feynman rules of the operator insertions. Furthermore, for exclusive processes the operator matrix elements are non-forward and mixing with total-derivative operators has to be taken into account, such that also the Feynman rules for total-derivative operator insertions are generically needed. We present our results in a form appropriate for implementation in a computer algebra system, and we provide sample implementations in {\tt Mathematica} and \textit{{\tt FORM}} which are available at \url{https://github.com/vtsam/NKLO}.

\subsection*{Acknowledgements}
We would like to thank J. Gracey for providing his private code for the $N=3$ Feynman rules for the purpose of cross-checking our results. Furthermore, we thank J. Gracey and S. Moch for valuable comments on the manuscript. This work has been supported by grant K143451 of the National Research, Development and Innovation Fund in Hungary. Furthermore, the work of G.S. was supported by the Bolyai Fellowship program of the Hungarian Academy of Sciences.

\bibliographystyle{JHEP}
\bibliography{omebib}

\end{document}